\documentclass[prl,twocolumn,showpacs]{revtex4}

\usepackage{color}
\usepackage{epsfig}
\usepackage{amsmath}
\usepackage{graphicx}

\newcommand{\sub}[1]{{\mbox{\scriptsize #1}}}

\begin{document}

\preprint{}

\title[]{Single-atom lasing induced atomic self trapping}

\author{Thomas Salzburger}
\author{Helmut Ritsch}
\affiliation{Institute for Theoretical Physics, University
Innsbruck, A 6020 Innsbruck, Austria}

\begin{abstract}
We study motion and field dynamics of a single-atom laser
consisting of a single incoherently pumped free atom moving in an
optical high-{\it Q} resonator. For sufficient pumping, the system
starts lasing whenever the atom is close to a field antinode. If
the field mode eigenfrequency is larger than the atomic transition
frequency, the generated laser light attracts the atom to the
field antinode and cools its motion. Using quantum Monte Carlo
wave function simulations, we investigate this coupled atom-field
dynamics including photon recoil and cavity decay. In the regime
of strong coupling, the generated field shows strong nonclassical
features like photon antibunching, and the atom is spatially
confined and cooled to sub-Doppler temperatures.
\end{abstract}


\maketitle

Optical cavity QED experiments using high finesse resonators and
ultracold neutral atoms have seen tremendous progress towards
larger coupling strength and interaction time in the past decade,
becoming a fruitful test ground of quantum theory
\cite{pinkse_hood,mckeever03a,iontrap,chapman03,photgun02}. One
particular goal of cavity QED is a single-atom, single-mode laser.
In analogy to the micromaser, a one atom laser using an atomic
beam crossing an optical resonator has been realized some time ago
\cite{an94}. However, although effectively only a single or a few
atoms are present at a time, the short interaction times make this
still a many particle device. Very recently important progress
towards single particle lasing has been achieved by placing a
small ion trap into a high-{\it Q} cavity \cite{iontrap,meyer97}.
This allows very long interaction times, but technical limits set
by the ion trap still prevent to reach threshold.

In all optical setups the atoms are extremely cold at the
beginning, but the light forces induced by the cavity field and
pumping itself lead to fast heating. Even less than a single
photon on average induces significant dissipation
\cite{pinkse_hood} and limits interaction times. By a proper
choice of parameters, one can minimize this heating caused by
light forces \cite{doherty01} or even use cavity induced cooling
forces \cite{domokos03,horak97} to enlarge the trapping times.
Alternatively, the addition of an extra dipole trapping potential
led to much larger interaction times \cite{mckeever03b} and
enabled the first realization of single atom lasing in the strong
coupling regime \cite{mckeever03a}. Here, steady state light
output generated from one atom for almost a second was achieved
from a three photon Raman type gain scheme.

In this Letter we go conceptually one step beyond and assume that
no coherent field is applied to the atom or the mode and no extra
atom trap is present. The cavity field is entirely generated by
gain from the incoherently pumped atom. Neglecting atomic motion,
such single atom laser models have been widely used in fundamental
studies of quantum laser theory \cite{lasertheory}. They
constitute light sources with unique properties. For example, one
is able to sustain a stationary single photon Fock state or
generate highly sub-Poissonian output \cite{photonoutput}. Note
that the same single atom is present during the whole time here.

Extending these models, we include the atomic motion governed by
the light forces of the laser field created by the atom itself. As
expected, the proper accounting of the cavity field forces implies
significant modifications of the system dynamics \cite{domokos03}.
In general, the atomic motional variables and the dynamics of the
internal variables are coupled and correlated or even entangled.
The atom moves under the influence of the dipole force which
depends on the mode intensity. This intensity in turn depends on
the atomic position in a highly nonlinear way as the atom itself
is the gain medium.

From a first guess, one could already expect some selftrapping
effect if one chooses the parameters such that lasing only starts
when the atom is close to a field antinode. The laser field then
could generate an attractive potential keeping the atom in the
vicinity of this antinode. Naively, this requires the lasing mode
to be red detuned relative to the atomic transition frequency. In
this case, though, the extra energy from the atomic transition is
likely to be converted into kinetic energy heating the atom.
Similarly, for a blue detuned cavity mode, one expects cooling but
repulsion from the antinodes. However, the situation is a bit more
complicated. As gain requires atomic inversion, an atom in steady
state can still be a high field seeker for blue detuning, and the
laser frequency is itself a dynamical quantity. Thus, we can still
hope to find parameters where trapping, cooling, and lasing
coincide. Of course, heating through spontaneous emission and
dipole fluctuations will also be present. This makes the total
dynamics hard to guess, which motivated us to study the problem in
more detail.

Let us now define our model as simple as possible still containing
the essential physics. For this we restrict ourselves to a
two-level atom moving in a single strongly coupled cavity mode in
one dimension. As in well proven approaches developed in the early
days of laser theory \cite{lasertheory}, incoherent pumping can be
consistently modelled by inverse spontaneous emission at rate
$2\delta$. Following standard procedures of quantum optics, we can
derive the master equation $\dot{\rho} = {\cal L}\rho$ governing
the time evolution of field and atom including atomic spontaneous
emission at rate $2\gamma$ and cavity decay at rate $2\kappa$.

In a frame rotating at the mode frequency $\omega_c$ we get:
\begin{widetext}
\begin{align}
   {\cal L}\rho &= i\Delta\Bigl[|e\rangle\langle e|,\rho\Bigr]_-
      - g \cos(kx)\left(\Bigl[|e\rangle\langle g|a,\rho\Bigr]_-
      -\Bigl[a^\dagger |g\rangle\langle e|,\rho\Bigr]_-\right)
      \nonumber \\
      &+ \delta\left(2|e\rangle\langle g|\rho|g\rangle\langle e| -
      \Bigl[|g\rangle\langle g|,\rho\Bigr]_+\right) +
      \gamma\left(2|g\rangle\langle e|\rho|e\rangle\langle g| -
      \Bigl[|e\rangle\langle e|,\rho\Bigr]_+\right) +
      \kappa\left(2a\rho a^\dagger - \Bigl[a^\dagger
      a,\rho\Bigr]_+\right).
\end{align}
\end{widetext}
Here $|g\rangle$ ($|e\rangle$) and $a$ denote the atomic ground
(excited) state and the field annihilation operator, respectively.
The cavity mode with mode function $\cos(kx)$ and frequency
$\omega_c$ is detuned from the atomic transition frequency
$\omega_a$ by $\Delta =\omega_c - \omega_a$.

\begin{figure}[b]
\begin{picture}(0,0)%
\includegraphics{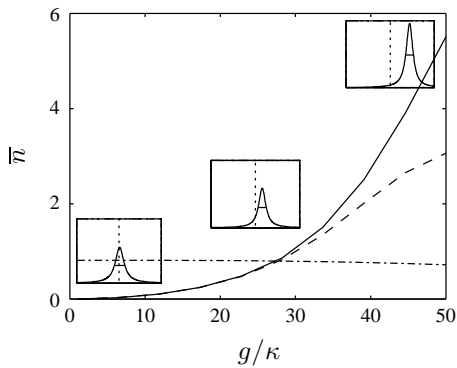}%
\end{picture}%
\setlength{\unitlength}{4144sp}%
\begingroup\makeatletter\ifx\SetFigFont\undefined%
\gdef\SetFigFont#1#2#3#4#5{%
  \reset@font\fontsize{#1}{#2pt}%
  \fontfamily{#3}\fontseries{#4}\fontshape{#5}%
  \selectfont}%
\fi\endgroup%
\begin{picture}(2758,2145)(-388,-1276)
\put(1006,-1231){\makebox(0,0)[lb]{\smash{\SetFigFont{10}{12.0}{\rmdefault}{\mddefault}{\updefault}{\color[rgb]{0,0,0}$g/\kappa$}%
}}}
\put(-269,-60){\rotatebox{90.0}{\makebox(0,0)[lb]{\smash{\SetFigFont{10}{12.0}{\rmdefault}{\mddefault}{\updefault}{\color[rgb]{0,0,0}$\overline{n}$}%
}}}}
\end{picture}
  \caption{Average photon number $\overline{n}$, scaled uncertainty
   $\Delta_n/\overline{n}$, and upper state population as function
   of $g/\kappa$. The parameters are $(\gamma,\delta,\Delta) =
   (10\kappa,75\kappa,250\kappa)$. The little inserts show the
   emitted light spectrum for $g=(5,25,45)\kappa$.}
   \label{fig:fig1}
\end{figure}

In a first step we look at the steady state of the system for
fixed atomic position $x$ which enters the equations only via the
coupling strength $g \cos(kx)$. In Fig.~\ref{fig:fig1} we plot the
photon number $\overline{n}$ (solid line), its scaled uncertainty
$\Delta_n/\overline{n}$ (dashed), and the atomic upper state
population (dash-dotted) as a function of $g$ for fixed pump
strength $\delta=75\kappa=7.5 \gamma$ for large atom-field
detuning $\Delta = 250 \kappa$. As expected, the photon number
depends upon $g$ in a nonlinear way and the system starts lasing
only for sufficiently large $g$ when the atom is close to a field
antinode. With growing photon number $(\overline{n}>1)$ the
corresponding spectrum shown in the little inserts of
Fig.~\ref{fig:fig1} is blue shifted from the atomic resonance and
acquires a width below the empty cavity linewidth. Hence, we can
hope for stable trapping close to the antinode combined with
lasing at these parameters.

Let us include atomic motion and check. Since the number of
degrees of freedom is rather large now, we use a Monte Carlo wave
function simulation technique to numerically approximate the
solution of the master equation \cite{montecarlo}. Averaging gives
an approximate density operator $\rho(t)$ while analyzing
individual trajectories provides extra insight in the
microscopical dynamics.

\begin{figure}[t]
\begin{picture}(0,0)%
\includegraphics{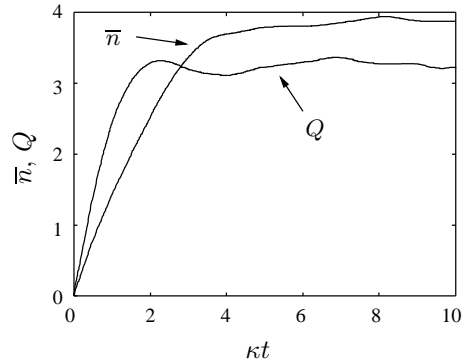}%
\end{picture}%
\setlength{\unitlength}{4144sp}%
\begingroup\makeatletter\ifx\SetFigFont\undefined%
\gdef\SetFigFont#1#2#3#4#5{%
  \reset@font\fontsize{#1}{#2pt}%
  \fontfamily{#3}\fontseries{#4}\fontshape{#5}%
  \selectfont}%
\fi\endgroup%
\begin{picture}(2751,2153)(-365,-1284)
\put(199,634){\makebox(0,0)[lb]{\smash{\SetFigFont{9}{10.8}{\rmdefault}{\mddefault}{\updefault}{\color[rgb]{0,0,0}$\overline{n}$}%
}}}
\put(1394, 92){\makebox(0,0)[lb]{\smash{\SetFigFont{9}{10.8}{\rmdefault}{\mddefault}{\updefault}{\color[rgb]{0,0,0}$Q$}%
}}}
\put(1036,-1239){\makebox(0,0)[lb]{\smash{\SetFigFont{10}{12.0}{\rmdefault}{\mddefault}{\updefault}{\color[rgb]{0,0,0}$\kappa t$}%
}}}
\put(-246,-196){\rotatebox{90.0}{\makebox(0,0)[lb]{\smash{\SetFigFont{10}{12.0}{\rmdefault}{\mddefault}{\updefault}{\color[rgb]{0,0,0}$\overline{n}$, $Q$}%
}}}}
\end{picture}
   \caption{Average photon number $\overline{n}$ and Mandel $Q$
     parameter for the lasing startup phase. The parameters are
     $(\gamma,\delta,g) = (10\kappa,60\kappa,50\kappa)$ where the
     cavity field is detuned from the atomic transition by
     $\Delta = 250\kappa$.}
   \label{fig:start}
\end{figure}

We briefly review the method now. Defining a superoperator ${\cal
S}\rho = \sum_i \hat{C}_i \rho \hat{C}_i^\dagger $ based on
collapse operators $\hat{C}_i$ related to the quantum jumps, we
get single trajectories by propagation with a non-hermitian
Hamiltonian $H\sub{eff}$
\begin{equation}
\label{eq:ls}
   ({\cal L-S})\rho = -i H\sub{eff}\rho + i \rho
   H\sub{eff}^\dagger,
\end{equation}
in between stochastically occurring jumps. The jump probabilities
within the interval $[t,t+\triangle t]$ are given by $p_i(t) =
\langle\psi(t)| \hat{C}_i^\dagger\hat{C}_i |\psi(t)\rangle
\triangle t$. One first propagates $|\psi(t)\rangle$ using
$H\sub{eff}$ for one time step $\triangle t$
\begin{equation}
   \label{eq:ev}
   |\psi(t+\triangle t)\rangle = \exp(-i H\sub{eff} \triangle
   t)|\psi(t)\rangle
\end{equation}
an calculates $p_i$. Using random numbers $r_i \in [0,1]$ and
comparing them to $p_i(t)$, one then decides on the occurrence of
collapses $|\psi\rangle \rightarrow \hat{C}_i |\psi\rangle$ at the
end of $\triangle t$. The jumps here include atomic decay
($\sqrt{2\gamma}|g\rangle\langle e|$), cavity decay
($\sqrt{2\kappa}a$), and pump events
($\sqrt{2\delta}|e\rangle\langle g|$).

Explicitly the effective Hamiltonian $H\sub{eff}$ reads $(\hbar =
1)$
\begin{align}
   H\sub{eff} = &-i\delta|g\rangle\langle g|
     - (\Delta+i\gamma)|e\rangle\langle e|
      - i\kappa a^\dagger a \nonumber \\
   &- i g\cos(kx)\left(|e\rangle\langle g|a -
      a^\dagger |g\rangle\langle e|\right) .
\end{align}

\begin{figure}[t]
\begin{picture}(0,0)%
\includegraphics{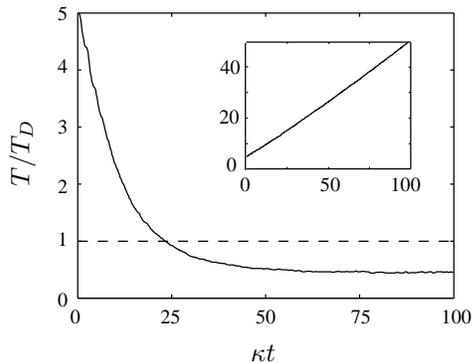}%
\end{picture}%
\setlength{\unitlength}{4144sp}%
\begingroup\makeatletter\ifx\SetFigFont\undefined%
\gdef\SetFigFont#1#2#3#4#5{%
  \reset@font\fontsize{#1}{#2pt}%
  \fontfamily{#3}\fontseries{#4}\fontshape{#5}%
  \selectfont}%
\fi\endgroup%
\begin{picture}(2839,2160)(-422,-1291)
\put(1036,-1246){\makebox(0,0)[lb]{\smash{\SetFigFont{10}{12.0}{\rmdefault}{\mddefault}{\updefault}{\color[rgb]{0,0,0}$\kappa t$}%
}}}
\put(-291,-196){\rotatebox{90.0}{\makebox(0,0)[lb]{\smash{\SetFigFont{10}{12.0}{\rmdefault}{\mddefault}{\updefault}{\color[rgb]{0,0,0}$T/T_D$}%
}}}}
\end{picture}
   \caption{Mean square velocity in units of the Doppler temperature
      $T_D$ for
      $\Delta = 250\kappa$. The other parameters are $(\gamma,\delta,g)
      = (10\kappa,60\kappa,50\kappa)$, $T_D$ is indicated by the dashed
      line. For negative detuning the atom gets heated up, see inset where
      $\Delta = -250\kappa$.}
   \label{fig:vel}
\end{figure}

As the atomic temperature is well above the recoil limit, we treat
the external atomic variables $x$ and $p$ classically
\cite{domokos01}. Thus we have $\dot{x} = p/m$ and $\dot{p} = F$
to add to our equations. The force $F$ is given by
\begin{equation}
   F =\langle \psi|-\frac{\partial H\sub{eff}}{\partial x}|\psi\rangle
      =-i\hbar k \sin kx \langle \psi|e\rangle\langle g|
      a|\psi\rangle+\mbox{h.c.}
 \end{equation}
Finally, we have to truncate the Hilbert space ${\cal H}
=\mbox{span}\{|g\rangle,|e\rangle\} \otimes{\cal H}_F$ (${\cal
H}_F$ is the Fock-space of the mode) at photon number $N$. Our
state space thus contains the ground state $|g,0\rangle$ and $N$
manifolds separated by $\omega_a$ with states $|g,n\rangle$ and
$|e,n-1\rangle$, $n = 1...N$. Any wave function is determined by
$2N+1$ complex numbers,
\begin{equation}
   |\psi\rangle = g_0|g,0\rangle +
   \sum_{n=1}^N\left(g_n|g,n\rangle + e_n|e,n-1\rangle\right) ,
\end{equation}
where only $2N$ are independent due to normalization. Note that in
contrast to most previous treatments of light forces in a cavity,
we cannot adiabatically eliminate the excited state since we have
to deal with an inverted atom.

Let us now examine stochastic simulation averages for typical
cases. After turning on the pump, the mean photon number
$\overline{n}$ increases rapidly and reaches a constant value
within a few cavity relaxation times (Fig.~\ref{fig:start}). At
the same time the atom gets trapped and remains captured in the
potential of the light it has generated. In addtition we show the
Mandel {\it Q} parameter ($ Q =
(\overline{n^2}-\overline{n}^2)/(\overline{n}) - 1 $) as a measure
of the field intensity noise.

\begin{figure}[t]
\begin{picture}(0,0)%
\includegraphics{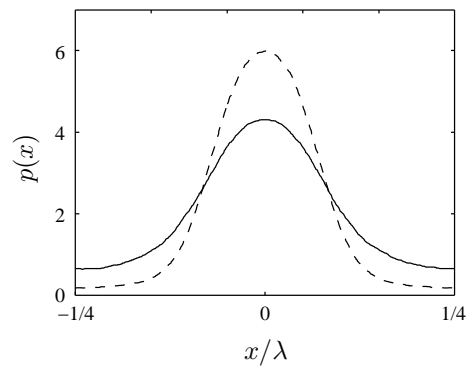}%
\end{picture}%
\setlength{\unitlength}{4144sp}%
\begingroup\makeatletter\ifx\SetFigFont\undefined%
\gdef\SetFigFont#1#2#3#4#5{%
  \reset@font\fontsize{#1}{#2pt}%
  \fontfamily{#3}\fontseries{#4}\fontshape{#5}%
  \selectfont}%
\fi\endgroup%
\begin{picture}(2745,2141)(-366,-1294)
\put(1016,-1254){\makebox(0,0)[lb]{\smash{\SetFigFont{10}{12.0}{\rmdefault}{\mddefault}{\updefault}{\color[rgb]{0,0,0}$x/\lambda$}%
}}}
\put(-247,-196){\rotatebox{90.0}{\makebox(0,0)[lb]{\smash{\SetFigFont{10}{12.0}{\rmdefault}{\mddefault}{\updefault}{\color[rgb]{0,0,0}$p(x)$}%
}}}}
\end{picture}
   \caption{Position distribution of an atom trapped at an antinode for
   the parameters of Fig.~\ref{fig:vel} (solid line) and for a larger
   pumping rate $\delta = 80\kappa$ (dashed line).}
   \label{fig:pos}
\end{figure}

Fig.~\ref{fig:vel} shows the atomic kinetic energy averaged over
4000 realizations in units of the Doppler temperature $T_D = \hbar
\gamma/2$ for the parameters of Fig.~\ref{fig:start} where the
mean photon number reaches $\overline{n} = 3.8$.  Starting at a
very low velocity, the atom gets cooled (heated) for a wide range
of positive (negative) detunings $\Delta$. The inset shows the
analogous result for $\Delta = -250\kappa$. This agrees with a
simple energy conservation argument. The atom gains the internal
energy $\hbar\omega_a$ from a pump event. Subsequently it loses
the energy $\hbar\omega_c$ by stimulated emission into the cavity
mode. The energy mismatch is compensated by the atomic center of
mass motion which gets damped if $\omega_c - \omega_a > 0$. This
argument is confirmed by a closer look at basic absorption and
emission processes including Doppler shift from the atomic motion.

Naively one would guess that the atom is expelled from the
interaction region of a blue shifted light field. However, this is
not true for a partially inverted atom which still can be trapped
near the mode antinodes. This feature is essential to allow steady
state operation of our single-atom laser. The solid line in
Fig.~\ref{fig:pos} shows the position distribution of an atom
trapped after the cooling process of Fig.~\ref{fig:vel}. Clearly,
the atom spends most of the time in the vicinity of an antinode
with a mean square distance of $\overline{x^2}=(0.1\lambda)^2$.
This value decreases further for larger photon numbers. For
$\overline{n} \approx 12$ (dashed line) we get
$\overline{x^2}=(0.07\lambda)^2$ for a pump strength of $\delta =
80\kappa$. The non-vanishing probability near $x\approx \pm
\lambda/4$ shows some remaining hopping of the atom between
different trapping sites.
\begin{figure}[t]
\begin{picture}(0,0)%
\includegraphics{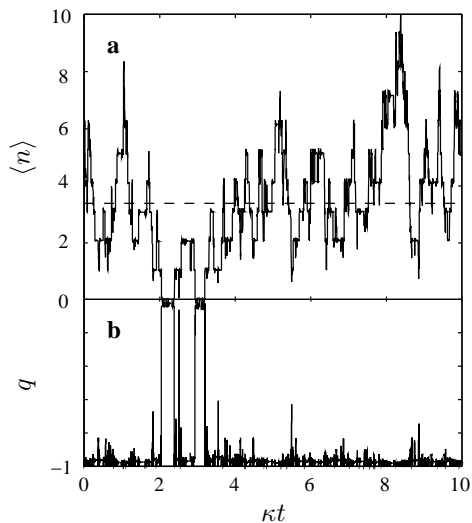}%
\end{picture}%
\setlength{\unitlength}{4144sp}%
\begingroup\makeatletter\ifx\SetFigFont\undefined%
\gdef\SetFigFont#1#2#3#4#5{%
  \reset@font\fontsize{#1}{#2pt}%
  \fontfamily{#3}\fontseries{#4}\fontshape{#5}%
  \selectfont}%
\fi\endgroup%
\begin{picture}(2814,3105)(451,-3136)
\put(1959,-3091){\makebox(0,0)[lb]{\smash{\SetFigFont{10}{12.0}{\rmdefault}{\mddefault}{\updefault}{\color[rgb]{0,0,0}$\kappa t$}%
}}}
\put(578,-2311){\rotatebox{90.0}{\makebox(0,0)[lb]{\smash{\SetFigFont{10}{12.0}{\rmdefault}{\mddefault}{\updefault}{\color[rgb]{0,0,0}$q$}%
}}}}
\put(571,-1021){\rotatebox{90.0}{\makebox(0,0)[lb]{\smash{\SetFigFont{10}{12.0}{\rmdefault}{\mddefault}{\updefault}{\color[rgb]{0,0,0}$\langle n\rangle$}%
}}}}
\end{picture}
   \caption{{\bf a} Photon number expectation value $\langle n\rangle$
     for a single trajectory of a trapped atom within ten cavity relaxation
     times. The parameters are as in Fig.~\ref{fig:vel}. The dashed
     line indicates the mean photon number.
     {\bf b} Corresponding single-trajectory $q$ factor.}
   \label{fig:single}
\end{figure}

\begin{figure}[bptm]
\begin{picture}(0,0)%
\includegraphics{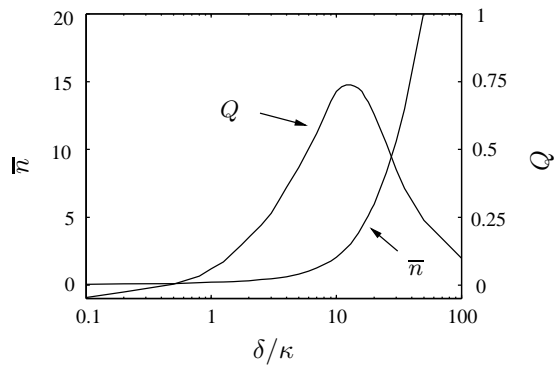}%
\end{picture}%
\setlength{\unitlength}{4144sp}%
\begingroup\makeatletter\ifx\SetFigFont\undefined%
\gdef\SetFigFont#1#2#3#4#5{%
  \reset@font\fontsize{#1}{#2pt}%
  \fontfamily{#3}\fontseries{#4}\fontshape{#5}%
  \selectfont}%
\fi\endgroup%
\begin{picture}(3276,2130)(-456,-1276)
\put(1936,-736){\makebox(0,0)[lb]{\smash{\SetFigFont{9}{10.8}{\rmdefault}{\mddefault}{\updefault}{\color[rgb]{0,0,0}$\overline{n}$}%
}}}
\put(811,209){\makebox(0,0)[lb]{\smash{\SetFigFont{9}{10.8}{\rmdefault}{\mddefault}{\updefault}{\color[rgb]{0,0,0}$Q$}%
}}}
\put(-337,-129){\rotatebox{90.0}{\makebox(0,0)[lb]{\smash{\SetFigFont{10}{12.0}{\rmdefault}{\mddefault}{\updefault}{\color[rgb]{0,0,0}$\overline{n}$}%
}}}}
\put(2776,-113){\rotatebox{90.0}{\makebox(0,0)[lb]{\smash{\SetFigFont{10}{12.0}{\rmdefault}{\mddefault}{\updefault}{\color[rgb]{0,0,0}$Q$}%
}}}}
\put(1005,-1231){\makebox(0,0)[lb]{\smash{\SetFigFont{10}{12.0}{\rmdefault}{\mddefault}{\updefault}{\color[rgb]{0,0,0}$\delta/\kappa$}%
}}}
\end{picture}
   \caption{Mean photon number and Mandel parameter
   vs pumping strength $\delta$.}
   \label{fig:scan}
\end{figure}

So far we calculated ensemble averages to investigate the motional
characteristics. It is now quite interesting to look at individual
trajectories  $|\psi(t)\rangle$ to visualize the microscopic
dynamics. Fig.~\ref{fig:single}a depicts a typical example of the
photon number time evolution $\langle n\rangle$ for a trapped atom
for ten cavity relaxation times. Starting in a certain state $|e,n
\rangle$, the system gets entangled with $|g,n+1\rangle$ owning to
the atom-cavity coupling. The incoherent pumping projects it into
the state $|e,n+1\rangle$ and creates a Fock state with $n+1$
photons. Unless a photon leaks out of the cavity or another one is
created, the photonic state remains nearly unchanged for the
following reasons. First, we chose a large detuning $\Delta =
250\kappa$ in order to keep the atom trapped at antinodes as good
as possible. Second, the atom undergoes several jumps due to
alternating spontaneous emission and pump events. After each of
these cycles the system is again projected into the initial state
$|e,n+1\rangle$, which gives the cascade shape of the photon
number time evolution. Notice that the pumping process creates no
coherence, as it would be e.g.~in a lambda type system such as in
\cite{photonoutput}. Here ac Stark splitting of dressed states
does not effect the probability of exciting the system which is
solely given by the atomic ground state population and, hence, the
system can be excited into states with higher photon numbers more
easily. Nevertheless, for each trajectory the cavity field remains
close to a Fock state with photon numbers varying around the mean
value indicated by the dashed line. This behavior is demonstrated
in Fig.~\ref{fig:single}b where the single trajectory Mandel
factor -- which describes the actual state of the cavity field --
attains values close to $q = -1$ as long as one photon is present
at least. This is due to the fact that in each trajectory we know
the number of pump and decay events. Statistical averaging over
the pump events washes out this feature.

Let us now come back to ensemble averages of field properties.
Fig.~\ref{fig:scan} depicts $\overline{n}$ and $Q$ as a function
of $\delta$ with $(\Delta,g) = (200\kappa,100\kappa)$ for $\gamma
= 0$. In general the field intensity is strongly fluctuating.
These fluctuations are even more pronounced than for an atom at
rest and in turn increase motional heating. Only for larger photon
numbers the Mandel $Q$ parameter drops down to zero as for a
coherent state (lasing).

In summary light forces significantly influence the dynamics of a
single-atom laser. Surprisingly, for blue atom-field detuning
several effects work together in a favorable way to facilitate
steady state lasing in conjunction with mechanical cooling and
trapping. Steady state temperatures below the Doppler limit are
possible in spite of enhanced momentum diffusion due to
fluctuations of the photon number. The light field shows
nonclassical features and approximates a coherent state with a
coherence time beyond the cavity lifetime only far above
threshold. Although the description of the atom is oversimplified
here, we believe that our central findings like enhanced trapping
and self cooling with lasing still should be present in a real
system.

This work was supported by the Austrian FWF under contract S1512.
The authors thank P.~Domokos and A.~Kuhn for helpful discussions.


\begin{thebibliography}{99}

\bibitem{pinkse_hood}
  P.~W.~H.~Pinkse {\it et al.}, Nature {\bf 404}, 365 (2000);
  C.~J.~Hood {\it et al.}, Science {\bf 287}, 1447 (2000).

\bibitem{iontrap}
  G.~R.~Guth\"orlein {\it et al.}, Nature {\bf 414}, 49 (2001);
  A.~B.~Mundt {\it et al.}, Phys.~Rev.~Lett.~{\bf 89}, 103001 (2002).

\bibitem{chapman03}
  J.~A.~Sauer {\it et al.}, quant-ph/0309052 (2002).

\bibitem{photgun02}
  A.~Kuhn, M.~Hennrich, and G.~Rempe, Phys.~Rev.~Lett.~{\bf 89}, 067901 (2002).

\bibitem{mckeever03a}
  J.~McKeever {\it et al.}, Nature {\bf 425}, 268 (2003).

\bibitem{an94}
  K.~An {\it et al}, Phys.~Rev.~Lett.~{\bf 73}, 3375 (1994).

\bibitem{meyer97}
  G.~M.~Meyer, H.-J.~Briegel, and H.~Walther,
  Europhys.~Lett.~{\bf 37}, 317 (1997).

\bibitem{doherty01}
  A.~C.~Doherty {\it et al.}, Phys.~Rev.~A {\bf 63}, 013401 (2001).

\bibitem{domokos03}
  P.~Domokos and H.~Ritsch,
  J.~Opt.~Soc.~Am.~B {\bf 20}, 1098 (2003).

\bibitem{horak97}
  P.~Horak {\it et al.}, Phys.~Rev.~Lett.~{\bf 79}, 4974 (1997).

\bibitem{mckeever03b}
  J.~McKeever {\it et al.}, Phys.~Rev.~Lett.~{\bf 90}, 133602 (2003).

\bibitem{lasertheory}
  H.~Haken, {\it Laser Theory} (Springer, Berlin, 1984);
  C.~Ginzel {\it et al.}, Phys.~Rev.~A {\bf 48}, 732 (1993).

\bibitem{photonoutput}
  T.~Pellizzari and H.~Ritsch, Phys.~Rev.~Lett.~{\bf 79}, 3973 (1994);
  H.~Ritsch {\it et al.}, Phys.~Rev.~A {\bf 44}, 3361 (1991).

\bibitem{domokos01}
  P.~Domokos, P.~Horak, and H.~Ritsch,
  J.~Phys.~B: At.~Mol.~Opt.~Phys.~{\bf 34}, 187 (2001).

\bibitem{montecarlo}
  H.~J.~Carmichael, {\it An Open System Approach to Quantum Optics}
  (Springer, Berlin, 1993);
  R.~Dum, P.~Zoller, and H.~Ritsch,
  Phys.~Rev.~A {\bf 45}, 4879 (1992);
  J.~Dalibard and Y.~Castin,
  Phys.~Rev.~Lett.~{\bf 68}, 580 (1992).


\end{thebibliography}
\end{document}